\documentclass[doublecol]{epl2}
\usepackage{amsmath}
\usepackage{graphicx}
\usepackage{dcolumn}
\usepackage{bm}
\usepackage{epsfig}
\usepackage[T1]{fontenc}
\usepackage{ae,aecompl}

\begin{document}

\title{Multiple-Relaxation-Time Lattice Boltzmann Approach to Compressible Flows
with Flexible Specific-Heat Ratio and Prandtl Number}
\shorttitle{MRT LB approach to compressible flows}
\author{Feng Chen\inst{1} \and Aiguo Xu\inst{2}\footnote{
Corresponding author. E-mail: Xu\_Aiguo@iapcm.ac.cn} \and Guangcai Zhang%
\inst{2} \and Yingjun Li\inst{1} \and Sauro Succi\inst{3}}
\shortauthor{Chen, Xu, Zhang, Li and Succi}

\institute{
  \inst{1} State Key Laboratory for GeoMechanics and Deep Underground Engineering, \\
  China University of Mining and Technology (Beijing), Beijing 100083, P.R.China \\
  \inst{2} National Key Laboratory of Computational Physics, \\
Institute of Applied Physics and Computational Mathematics, P. O.
Box 8009-26, Beijing 100088, P.R.China \\
  \inst{3} Istituto Applicazioni Calcolo-CNR, Viale del Policlinico 137, 00161,
Roma, Italy}


\abstract{A new multiple-relaxation-time lattice Boltzmann scheme
for compressible flows with arbitrary specific heat ratio and
Prandtl number is presented. In the new scheme, which is based on a
two-dimensional $16$-discrete-velocity model, the kinetic moment
space and the corresponding transformation matrix are constructed
according to the seven-moment relations associated with the local
equilibrium distribution function. In the continuum limit, the model
recovers the compressible Navier-Stokes equations with flexible
specific-heat ratio and Prandtl number. Numerical experiments show
that compressible flows with strong shocks can be simulated by the
present model up to Mach numbers $Ma \sim 5$.}

\maketitle

\section{Introduction}

Over the last two decades, the Lattice Boltzmann (LB) method has become a
prominent tool in computational fluid dynamics\cite{2,BSV,AIDUN}, with a
broad spectrum of complex flows applications, ranging from multiphase flows%
\cite{ShanChen,Swift,XGL1}, magnetohydrodynamics\cite%
{PRL1991,PRA1991,PRE2002,PRE2004}, flows through porous media\cite%
{10,10a,CALI} and thermal fluid dynamics\cite{11,12}, to cite but a few.
Being based on a low-Mach expansion of Maxwell-Boltzmann local equilibria,
standard LB models are typically used for the simulation of
quasi-incompressible flows. Given the great importance of shock-wave
dynamics in many fields of physics and engineering \cite%
{SHOCK1,materialpoint}, constructing LB models for high-speed compressible
flows with shocks has been attempted since the early days of LB research.
However, current LB versions for compressible flows are still subject to at
least one of the following constraints: low Mach number, fixed Prandtl
number and fixed specific heat ratio. Broadly speaking, to date, there are
two LB approaches for the simulation of compressible flows. The first is
based on the Single-Relaxation-Time (SRT) Bhatnagar-Gross-Krook (BGK)
approximation\cite{5}, where the local equilibrium distribution function $%
f_{i}^{eq}$ is calculated from a truncated Taylor expansion of the flow
velocity, where $i$ is the index of discrete velocity. Models along this
line can be roughly categorized into two groups, the standard LB and the
Finite Difference (FD) LB. Examples in the first group are referred to the
works of Alexander and Chen et al.\cite{Alexander1}, Sun, et al.\cite{sun1},
Yan, et al. \cite{yan1}, and so on. In the FDLB formulation, space and time
discretizations are independent, and to overcome the limit of low Mach
number, one has to confine the LB to be a solver of Euler or Navier-Stokes
(NS) equations. In such a case, the discrete $f_{i}^{eq}$ is best seen as a
local attractor of the collisional relaxation process, leading to the
desired form of the macroscopic moments. Examples are referred to the works
of Tsutahara, et al\cite{20b,21a}, Xu, et al\cite{43}, Shu, et al\cite%
{CShuPRE} and He, et al\cite{hewl}. The application of BGK also
leads to a fixed Prandtl number. In most existing models, the
specific heat ratio is fixed to a nonrealistic constant because only
the translational degrees of freedom are taken into account in the
internal energy. The second way consists of resorting to the
original scattering-matrix formulation of the LB equation\cite{HSB},
whose optimized version is nowadays known as
Multiple-Relaxation-Time (MRT) LB scheme\cite{13}. In the MRT-LB
formulation, the collision step is first calculated in the kinetic
moment space (KMS), i.e. the space spanned by the kinetic moments
defined by the projection of the distribution function onto a basis
set of $N$ Hermite polynomials. Subsequently, the streaming step is
performed back in the discrete velocity space spanned by a suitable
set of $N$ discrete velocities $\mathbf{v}_i$.
In contrast to the SRT model,
the MRT version caters for more adjustable parameters and degrees of
freedom. The relaxation rates of the various kinetic moments due to
particle collisions may be adjusted independently. This overcomes
some obvious deficiencies of the SRT model, such as a fixed Prandtl
number. The low-Mach number constraint is generally caused by the
numerical instability problem. To this regard, extensive efforts
have been made in the past years, such as the entropic
method\cite{16+,17+,ENT}, the fix-up scheme\cite{16+,18+},
Flux-limiters\cite{Sofonea1} and dissipation\cite{43,Brownlee1}
techniques. In many cases with low-speed flows, it has been shown
\cite{17} that the MRT LB model also offers enhanced numerical
stability. As far as we know, nearly all existing MRT models have
been focussing on the standard LB for isothermal systems without
strong shocks.

Recently, we presented a thermal MRT FDLB model for high-speed
compressible flows\cite{submitted}. The MRT model uses $16$ discrete
velocities, as suggested by Kataoka and Tsutahara\cite{21a} in the
SRT version. In that MRT model, the KMS and corresponding
transformation matrix are constructed according to the group
representation theory; equilibria of the nonconserved moments are
chosen so as to recover the compressible Navier-Stokes equations
through the Chapman-Enskog analysis. Neither the formulation, nor
the simulation procedure are by any means related to a
truncated Taylor-expansion of local equilibrium distribution function $%
f_{i}^{eq}$. Consequently, the discrete local equilibrium $f_{i}^{eq}$
remains Maxwellian regardless of the value of the Mach number. However, the
model is restricted to a nonphysical value of the specific-heat-ratio, $2$.
For convenience of description, we will refer to that model as model $I$.

In this letter, we formulate a LB model II for compressible flows
with arbitrary specific-heat-ratio. At variance with model I, at
high-Mach number, the local equilibria of model II depart from a
Maxwellian. Nevertheless, the method is able to simulate both low
and high Mach numbers regimes (up to $Ma \sim 5$).

\section{ Brief review of the MRT LB model}

According to the main strategy of the MRT-LB scheme, the differential form
of the MRT-LB equation read as follows:
\begin{equation}
\frac{\partial f_{i}}{\partial t}+v_{i\alpha }\frac{\partial f_{i}}{\partial
x_{\alpha }}=-\mathbf{M}_{il}^{-1}\hat{\mathbf{S}}_{lk}(\hat{f}_{k}-\hat{f}%
_{k}^{eq})\text{,}  \label{3}
\end{equation}%
where $\mathbf{v}_{i}$ is the discrete particle velocity, $i=1$,$\ldots$ ,$N$, $N$\ is the
number of discrete velocities, the subscript $\alpha $\ indicates $x$\ or $y$%
. The variable $t$ is time, $x_{\alpha }$ is the spatial coordinate.
The matrix
$\hat{\mathbf{S}}=\mathbf{MSM}^{-1}=diag(s_{1},s_{2},\cdots ,s_{N})$
is the diagonal relaxation matrix, $f_{i}$ and $\hat{f}_{i}$ are the
particle distribution function in the velocity space and the kinetic
moment space respectively, $\hat{f}_{i}=m_{ij}f_{j}$, $m_{ij}$ is an
element of the matrix $\mathbf{M}$. Obviously, the mapping between
KMS and
velocity space is defined by the linear transformation $\mathbf{M}$, i.e., $%
\hat{\mathbf{f}}=\mathbf{Mf}$, $\mathbf{f=M}^{-1}\hat{\mathbf{f}}$, where
the bold-face symbols denote N-dimensional column vectors, e.g., $\mathbf{f}%
=( f_{1},f_{2},\cdots ,f_{N})^{T}$, $\hat{\mathbf{f}}=( \hat{f}_{1},\hat{f}%
_{2},\cdots ,\hat{f}_{N})^{T}$, $\mathbf{M}=( m_{1},m_{2},\cdots ,m_{N})
^{T} $, $m_{i}=(m_{i1},m_{i2},\cdots ,m_{iN})$. $\hat{f}_{i}^{eq}$\ is the
equilibrium value of the moment $\hat{f}_{i}$. The equation reduces to the
usual lattice BGK equation if all the relaxation parameters are set to be a
single relaxation time $\tau $, namely $\hat{\mathbf{S}}=\frac{1}{\tau }%
\mathbf{I}$, where $\mathbf{I}$ is the identity matrix.

\section{Construction of energy-conserving MRT LB model}

We use the two-dimensional discrete velocity model by Kataoka and Tsutahara
(see Fig. 1), which reads as follows:
\begin{equation*}
( v_{ix,}v_{iy}) =\left\{
\begin{array}{cc}
\mathbf{cyc}:(\pm 1,0) , & \text{for }1\leq i\leq 4, \\
\mathbf{cyc}:(\pm 6,0) , & \text{for }5\leq i\leq 8, \\
\sqrt{2}(\pm 1,\pm 1) , & \text{for }9\leq i\leq 12, \\
\frac{3}{\sqrt{2}}(\pm 1,\pm 1) , & \text{for }13\leq i\leq 16,%
\end{array}%
\right.
\end{equation*}%
where \textbf{cyc} indicates the cyclic permutation. The $16$ velocities are
grouped into four energy levels, $v^{2}$, where $v=1,2,3,6$. In this model,
besides the translational degrees of freedom, a parameter $\eta _{i}$ is
introduced, in order to describe the $(b-2)$ extra-degrees of freedom
corresponding to molecular rotation and/or vibration, where $\eta _{i}=5/2$
for $i=1$, $\cdots $,$4$, and $\eta _{i}=0$ for $i=5$, $\cdots $, $16$.
Details are referred to the original publication\cite{21a}. In our
simulations, the time evolution is based on the usual first-order upwind
scheme, while space discretization is performed through a Lax-Wendroff
scheme. 
\begin{figure}[tbp]
{%
\centerline{\epsfig{file=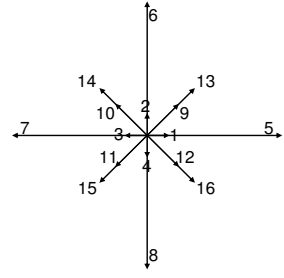,bbllx=27pt,bblly=61pt,bburx=315pt,bbury=349pt,
width=0.20\textwidth,clip=}}} \caption{ Distribution of
$\mathbf{v}_{i}$ for the discrete velocity model.}
\end{figure}

The moment representation provides a convenient way to express collisional
relaxation on different time scales for the various kinetic moments.
However, the choice of the kinetic moments is not unique. They can be
identified with physical quantities, such as density, momentum, energy,
momentum and energy fluxes, and so on. Inspired by a close tie with tensor
Hermite polynomials, many researchers choose the moments according to the
monomials of Cartesian components of the discrete velocities $%
v_{ix}^{m}v_{iy}^{n}$, $m,n=0,1,2,\cdots $\cite{17}. In this Letter,
the KMS and equilibria of the moments are chosen according to the
seven-moment relations (Equations (5a-5g) in Reference \cite{21a}),
associated with the local equilibrium distribution function $f_{i}^{eq}$%
(Equation (8) in Reference \cite{21a}), which is a polynomial of the flow
velocity up to the third order. The incorporation of the parameter $b$
permits to tune the specific-heat-ratio, $\gamma = (b+2)/b$.

The right-hand-sides of the seven equations indicate seven monomials, $1$, $%
v_{i\alpha }$, $(v_{i\alpha }^{2}+\eta _{i}^{2})$, $v_{i\alpha }v_{i\beta }$%
, $(v_{i\beta }^{2}+\eta _{i}^{2})v_{i\alpha }$, $v_{i\alpha }v_{i\beta
}v_{i\chi }$, $(v_{i\chi }^{2}+\eta _{i}^{2})v_{i\alpha }v_{i\beta }$. In
order to make the choice of moments more transparent, we construct the
transformation matrix through a simple combination of the seven monomials
above (see Appendix for details). As a result, at least eight of these
moments carry an explicit physical meaning: $\hat{f}_{1}=m_{1i}f_{i}$ is the
fluid density; $\hat{f}_{2}=m_{2i}f_{i}$ and $\hat{f}_{3}=m_{3i}f_{i}$
correspond to the $x-$ and $y-$ components of momentum (mass flux); $\hat{f}%
_{4}=m_{4i}f_{i}$ corresponds to the total energy mode; $\hat{f}%
_{6}=m_{6i}f_{i}$ and $\hat{f}_{7}=m_{7i}f_{i}$ correspond to the diagonal
and off-diagonal components of the stress tensor; $\hat{f}_{8}=m_{8i}f_{i}$
and $\hat{f}_{9}=m_{9i}f_{i}$ correspond to the $x-$ and $y-$ components of
the total energy flux.

At variance with previous MRT models\cite{17,17a}, the transformation matrix
$\mathbf{M}$ used in this work is not based upon a Gram-Schmidt
orthogonalization procedure. This is because such a procedure does not lead
to a conserved energy, as required for a consistent thermal and compressible
model. The moments can be divided into two groups. The first group consists
of those which are locally conserved in the collision process, i.e. $\hat{f}%
_{i}=\hat{f}_{i}^{eq}$. There are four such conserved moments,
density $\rho $, momenta $j_{x}$, $j_{y}$, and energy $e$ in two
dimensions, which are denoted by $\hat{f}_{1}$, $\hat{f}_{2}$,
$\hat{f}_{3}$, $\hat{f}_{4}$ mentioned above, respectively. The
second group consists of the non-conserved ones, i.e.
$\hat{f}_{i}\neq \hat{f}_{i}^{eq}$. The equilibria of the
non-conserved moments are functionals of the conserved ones. For the
collision process, one has maximum freedom in the construction of
the equilibrium functions of the non-conserved moments, provided the
basic principles of physics are satisfied (conservations of mass,
momentum and energy). In this model, the corresponding equilibrium
distribution functions in the KMS $\hat{f}_{i}^{eq}$ are chosen
according to the seven-moment relations, too (see Appendix). By
using the Chapman-Enskog expansion\cite{22a} on the two sides of the
LB equation, the final NS equations for compressible fluids read as
follows:
\begin{subequations}
\begin{equation}
\frac{\partial \rho }{\partial t}+\frac{\partial j_{x}}{\partial x}+\frac{%
\partial j_{y}}{\partial y}=0,  \label{12a}
\end{equation}%
\begin{eqnarray}
&&\frac{\partial j_{x}}{\partial t}+\frac{\partial }{\partial x}\left( \frac{%
j_{x}^{2}}{\rho }\right) +\frac{\partial P}{\partial x}+\frac{\partial }{%
\partial y}\left( \frac{j_{x}j_{y}}{\rho }\right)   \notag \\
&=&\frac{\partial }{\partial x}[\rho RT(\frac{(b-2)D}{bs_{5}}+\frac{d}{s_{6}}%
)]+\frac{\partial }{\partial y}\frac{\rho RTs_{xy}}{s_{7}}\text{,}
\label{12b}
\end{eqnarray}%
\begin{eqnarray}
&&\frac{\partial j_{y}}{\partial t}+\frac{\partial }{\partial x}\left( \frac{%
j_{x}j_{y}}{\rho }\right) +\frac{\partial }{\partial y}\left( \frac{j_{y}^{2}%
}{\rho }\right) +\frac{\partial P}{\partial y}  \notag \\
&=&\frac{\partial }{\partial y}[\rho RT(\frac{(b-2)D}{bs_{5}}-\frac{d}{s_{6}}%
)]+\frac{\partial }{\partial x}\frac{\rho RTs_{xy}}{s_{7}}\text{,}
\label{12c}
\end{eqnarray}%
\begin{eqnarray}
&&\frac{\partial e}{\partial t}+\frac{\partial }{\partial x}%
[(e+2P)j_{x}/\rho ]+\frac{\partial }{\partial y}[(e+2P)j_{y}/\rho ]  \notag
\\
&=&2\frac{\partial }{\partial x}\{\frac{\rho RT}{s_{8}}[(\frac{b}{2}+1)R%
\frac{\partial T}{\partial x}+s_{xy}u_{y}+(2\frac{\partial u_{x}}{\partial x}%
-\frac{2}{b}D)u_{x}]\}  \notag \\
&&+2\frac{\partial }{\partial y}\{\frac{\rho RT}{s_{9}}[(\frac{b}{2}+1)R%
\frac{\partial T}{\partial y}+s_{xy}u_{x}  \notag \\
&&+(2\frac{\partial u_{y}}{\partial y}-\frac{2}{b}D)u_{y}]\}\text{,}
\label{12d}
\end{eqnarray}%
where $P=\rho RT$, $e=b\rho RT+j_{\alpha }^{2}/\rho $ is twice the
total energy, $s_{xy}\equiv (\partial _{x}u_{y}+\partial
_{y}u_{x})$, $D\equiv (\partial _{x}u_{x}+\partial _{y}u_{y})$, and
$d\equiv (\partial _{x}u_{x}-\partial _{y}u_{y})$,
$s_{5},s_{6},s_{7}$ are related to viscosity, $s_{8},s_{9}$ are
related to heat conductivity. The relaxation parameters are not
completely independent, since the isotropy constraints impose
coupling between some of
these relaxation parameters, e.g., $s_{8}=s_{9}$. Since the elements of $%
\hat{\mathbf{S}}$ represent the inverse of the relaxation time for $\hat{%
\mathbf{f}}$ to its equilibrium $\hat{\mathbf{f}}^{eq}$ in KMS, and
the values of $\rho $, $j_{x}$, $j_{y}$, and $e$ are conserved in
the relaxation process, the values of $s_{1},s_{2},s_{3},s_{4}$ can
be fine-tuned independently. The remaining relaxation parameters
must belong to
the interval $[0,1/dt]$ for reasons of linear stability. In the limit $%
s_{5}=s_{6}=s_{7}=s_{8}=s_{9}=s$, the above NS equations reduce to the SRT
compressible NS equations.

\section{Numerical Simulation and Analysis}

In this section we study the following problems using the new MRT LB model:
measure of the sound speed, one- and two-dimensional Riemann problems,
Richtmyer-Meshkov instability. We work in a frame where the constant $R=1$.

\textit{(a) Speed of sound }

In order to validate the proposed model, the sound speed is
calculated and compared to the theoretical value $c_{s}=\sqrt{\gamma
T}$ \cite{20b}. A one-dimensional tube is used. In the tube, a plate
divides the fluids,
that share the same temperature and feature a small difference in pressure ($%
p_{1}/p_{2}=1.0 + 10^{-10}$) at the initial stage. The pressure ratio is so
small that expansion and compression waves propagate at the sound speed in
both directions when the plate is removed. The position of the pressure jump
is measured to calculate the speed of sound. The results at various
temperature levels are shown in Fig. 2, and the calculated values are found
to agree well with the theoretical ones.
\begin{figure}[tbp]
{%
\centerline{\epsfig{file=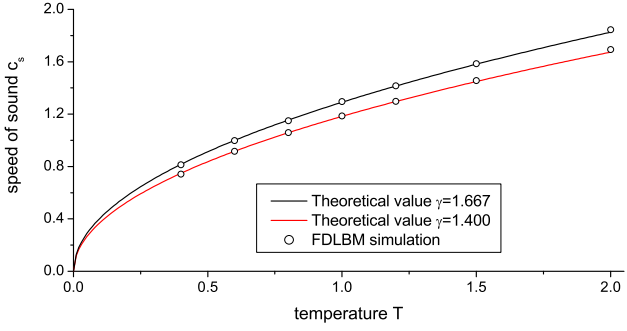,bbllx=12pt,bblly=11pt,bburx=287pt,bbury=163pt,
width=0.48\textwidth,clip=}}}
\caption{Calculation results and theoretical values of sound speed.}
\end{figure}

\textit{(b) One-dimensional Riemann problem}

Here, the two-dimensional model is used to solve the one-dimensional Riemann
problem, so as to visually show the advantages of this model as compared to
its SRT version. Consider the Riemann problem with initial condition given
by
\end{subequations}
\begin{equation*}
\begin{array}{cc}
(\rho ,u_{1},u_{2},T)|_{L}=(5.99924,19.5975,0.0,76.8254)\text{,} &  \\
(\rho ,u_{1},u_{2},T)|_{R}=(5.99242,-6.19633,0.0,7.69222)\text{,} &
\end{array}%
\end{equation*}
where subscript \textquotedblleft L\textquotedblright\ and \textquotedblleft
R\textquotedblright\ indicate the left and right macroscopic variables of
discontinuity. The numerical and exact solutions for two different
specific-heat ratios $\gamma =7/5$ and $\gamma =5/3$ at time $t=0.13$ are
shown in Fig.3, where the common parameters for SRT and MRT simulations are $%
dx=dy=0.002$, $dt=10^{-6}$. The relaxation time in SRT is $\tau =10^{-5}$,
while the collision parameters in MRT are $s_{5}=5\times 10^{3}$, $%
s_{6}=10^{4}$, and other values of $s$ are $10^{5}$. In the $x$ direction, $%
f_{i}=f_{i}^{eq}$ is set on the boundary nodes before the disturbance
reaches the two ends. The equilibrium distribution function $f_{i}^{eq}$ in
velocity space can be obtained via the linear transformation, $\mathbf{%
f^{eq}=M}^{-1}\hat{\mathbf{f}}^{eq}$. Along the $y$ direction,
periodic boundary conditions are adopted. The red triangle and red
circle symbols correspond to MRT simulation results with $\gamma
=7/5$ and $\gamma =5/3$, respectively; the green triangle and green
circle symbols correspond to SRT simulation results with $\gamma
=7/5$ and $\gamma =5/3$, respectively; the solid lines represent the
exact solutions for the case $\gamma =7/5$, and the dashed lines
represent the exact solutions for $\gamma =5/3$. We find that the
oscillations at the discontinuity are weaker in MRT simulations than
in their SRT countrepart. This shows that the problematic
``wall-heating" phenomenon is weaker in the MRT model than in its
SRT
version. 
\begin{figure}[tbp]
{%
\centerline{\epsfig{file=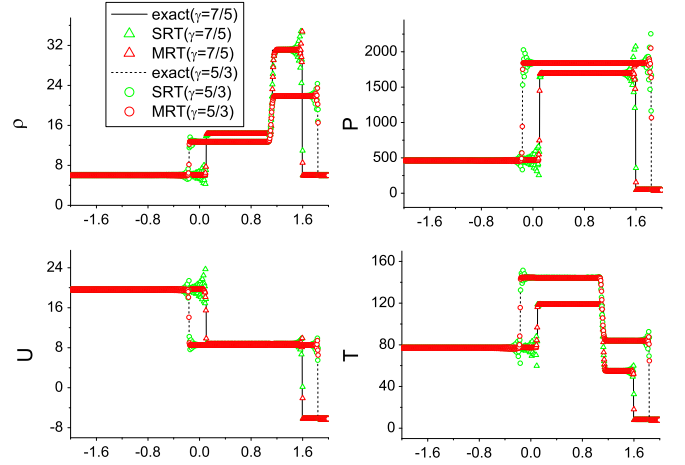,bbllx=28pt,bblly=24pt,bburx=422pt,bbury=310pt,
width=0.48\textwidth,clip=}}}
\caption{The numerical and exact solutions for the one-dimensional Riemann
problem at time $t=0.13$.}
\end{figure}

\textit{(c) Two-dimensional Riemann problem }

Compared with the relatively simple 1-D configurations, the 2-D Riemann
problem consists of a plethora of geometric wave patterns that pose a
challenge for the simulation. In two-dimensions, the Riemann problem
consists of four uniform states in each quadrant, $(\rho
,u_{1},u_{2},T)(x,y,0)=(\rho_{i},u_{1i},u_{2i},T_{i}), i=1,2,3,4$, where $i$
denotes the $i$th quadrant. We solve the 2-D Riemann problem with initial
data as follows:
\begin{equation*}
\begin{array}{c}
\rho _{1}=1.5\text{,}u_{11}=0\text{,}u_{21}=0\text{,}T_{1}=1\text{;} \\
\rho _{2}=0.5323\text{,}u_{12}=1.206\text{,}u_{22}=0\text{,}T_{2}=0.3/0.5323%
\text{;} \\
\rho _{3}=0.138\text{,}u_{13}=1.206\text{,}u_{23}=1.206\text{,}%
T_{3}=0.029/0.138\text{;} \\
\rho _{4}=0.5323\text{,}u_{14}=0\text{,}u_{24}=1.206\text{,}T_{4}=0.3/0.5323%
\text{.}%
\end{array}%
\end{equation*}
Figure 4 shows the density contours of the 2-D Riemann problem by MRT LB
simulation at time $t=0.12$, where the parameters are $dx=dy=0.002$, $%
dt=10^{-5}$, $s_{8}=s_{9}=2\times 10^{4}$, other values of $s$\ are
$3\times 10^{3}$. The Mach number of the second and fourth quadrants
are $1.358$, and the third quadrant is $3.1444$. While the SRT
version is found to fail, the MRT LB successfully recovers the main
features observed in earlier computations \cite{2dvp2,2dvp3}.
\begin{figure}[tbp]
{%
\centerline{\epsfig{file=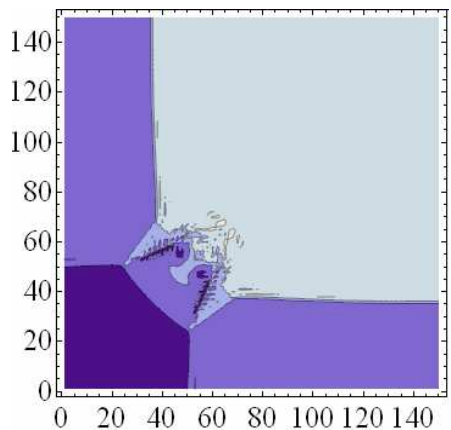,bbllx=68pt,bblly=180pt,bburx=339pt,bbury=451pt,
width=0.32\textwidth,clip=}}}
\caption{(Color online) Density contour lines of the 2-D Riemann problem in
MRT LB simulation at time $t=0.12$. From purple to gray, the density value
increases. }
\end{figure}

\textit{(d) Richtmyer - Meshkov instability}

Studies of the shock-induced Richtmyer - Meshkov (RM) instability are of
great importance from both viewpoints of fundamental research and practical
applications. For example, the RM instability occurs in Inertial Confined
Fusion, Scramjet engine, supersonic and hypersonic combustion, and others.
In addition, it also plays an important role in some natural phenomena, such
as Supernova explosions, formation of salt domains, and others. Although
most practical problems are three-dimensional, 2D studies are also of
recognized value, as they offer useful insights into the basic physics of
these problems. The simulation of RM instability still remains a challenging
topic to this day.

The initial configuration of our simulation is illustrated in Fig. 5(a),
where an interface with sinusoidal perturbation separates two different
fluids and a incident shock wave, with the Mach number $2.5$, traveling from
the right side, hits the interface immediately. The computational domain is
a rectangle with length $0.36$ and height $0.1$, which is divided into $%
360\times100$ mesh-cells. The initial sinusoidal perturbation at the
interface is: $x=0.24+0.005\times \sin(\pi/2+40\pi y)$, where the cycle of
initial perturbation is $0.05$, and the amplitude is $0.005$. The initial
conditions are as follows:
\begin{equation*}
\left\{
\begin{array}{cc}
(\rho,u_{1},u_{2},p)_{l}=(0.1358,0,0,1) \text{,} &  \\
(\rho,u_{1},u_{2},p)_{m}=(1,0,0,1) \text{,} &  \\
(\rho,u_{1},u_{2},p)_{r}=(3.33333,-2.07063,0,7.125) \text{,} &
\end{array}%
\right.
\end{equation*}%
where the subscripts $l$, $m$, $r$ indicate the left, middle, right regions
of the whole domain. The following boundary conditions are imposed: (1)
inflow at the right side; (2) reflecting condition at the left boundary, and
(3) periodic boundary conditions are applied at the top and bottom
boundaries. The reflecting boundary condition implies that the $x$ component
of the fluid velocity on the boundary is the reverse of the one of the
mirror-image interior point. The parameters are as follows: $\gamma =1.4$, $%
dt=10^{-5}$, $s_{5}=2\times 10^{3}$, and $10^{5}$ for the others.

When the shock wave passes the interface from the right, a reflected
rarefaction wave to the right and a transmission wave to the left,
are generated, and the perturbation amplitude decreases with the
interface motion to the left (Figure 5 (b)). Then, the peak and
valley of initial interface invert, the heavy and light fluids
gradually penetrate into each other as time goes on, the light fluid
``rises" to form a bubble and the heavy fluid ``falls" to generate a
spike ( shown in Figure 5 (c)). The transmission wave reaches the
solid wall on the left and reflects to the right, encounters the
interface again, converts into a transmission wave penetrating into
the heavy fluid zone and a reflection wave back to the light fluid.
Before the second shock wave reaches the interface, the interface
continues to move, and the disturbance develops at a lower growth
rate. When the second shock reaches the interface, the instability
growth rate of the disturbance significantly enhances. The
disturbance of the interface continues to grow, eventually forming a
mushroom shape (Figure 5 (d)). The simulation results show a
satisfactory agreement with those of other numerical simulations
\cite{rm1,rm2} and experiments\cite{rm4,rm5}. In the experimental
study of Puranik et al.\cite{rm4} the shock wave travels from light
(air) to heavy ($CO_{2}$) medium, while the reverse is true in our
simulation. As a result, interface reversals are observed in our
simulation. The same mechanism is also mentioned in Ref. \cite{rm4}
(see Fig. 11 in \cite{rm4}), which is a typical characteristic of an
incident shock from the heavier to the lighter material. Even though
the region with light medium is a cylindrical bubble in their
experiments, while it is a rectangle with distortions in our
simulation, the experiments on the interaction of a shock wave with
a single cylindrical bubble \cite{rm5} also show high similarity
with our simulation results, both in terms of the reversal mechanism
and of the mushroom shape.
\begin{figure}[tbp]
{%
\centerline{\epsfig{file=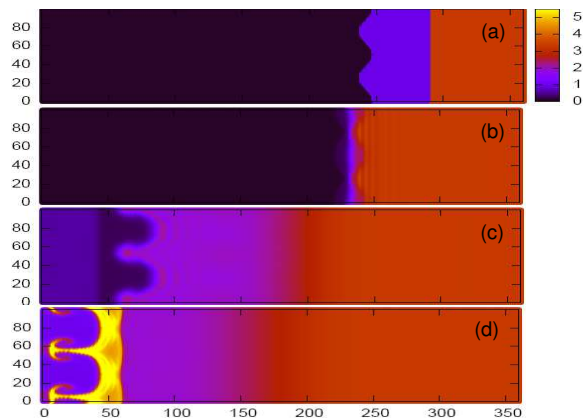,bbllx=27pt,bblly=61pt,bburx=539pt,bbury=429pt,
width=0.42\textwidth,clip=}}}
\caption{ (Color online) Snapshots of RM instability: density contour lines
at times $t=0$, $t=0.02$, $t=0.074$, $t=0.104$, respectively. From blue to
red the color corresponds to the increasing of density.}
\label{fig:wide}
\end{figure}

\section{Conclusions and remarks}

In this Letter, we have proposed a MRT-FDLB model for compressible flows
with flexible specific heat ratio and Prandtl number, as appropriate for
most applications. Different transport coefficients, such as viscosity and
heat conductivity, are related to different collision parameters, thereby
allowing a separate control of the different transport processes. The new
scheme has been validated for a series of one and two-dimensional numerical
benchmarks, always showing satisfactory agreement with theoretical results
and previous numerical work. Three-dimensional extensions of the present
scheme appear conceptually straightforward and shall make the object of
future work.

A. Xu and G. Zhang acknowledge support of the Science Foundations of LCP and
CAEP[under Grant Nos. 2009A0102005, 2009B0101012], National Natural Science
Foundation of China [under Grant Nos. 10775018, 10702010]. F. Chen and Y. Li
acknowledge support of National Basic Research Program of China [under Grant
No. 2007CB815105].

\section{Appendix. Construction of the transformation matrix and $\hat{f}%
_{i}^{eq}$}

In order to construct the transformation matrix in KMS, we take the
monomial $v_{i\alpha }v_{i\beta }$ as an example. Three
possibilities arise: (a) $\alpha =\beta =x$, $v_{i\alpha }v_{i\beta
}=v_{ix}^{2}$, (b) $\alpha =\beta =y $, $v_{i\alpha }v_{i\beta }=v_{iy}^{2}$%
, (c) $\alpha =x$, $\beta =y$, $v_{i\alpha }v_{i\beta }=v_{ix}v_{iy}$.
``(a)+(b)" gives $(v_{ix}^{2}+v_{iy}^{2})$, ``(a)-(b)" gives $%
(v_{ix}^{2}-v_{iy}^{2})$. In this way, the transformation matrix can be
composed as follows: $\mathbf{M}=(m_{1},m_{2},\cdots ,m_{16})^{T}$, where $%
m_{1i}=1\text{,} \qquad m_{2i}=v_{ix}\text{,} \qquad m_{3i}=v_{iy}\text{,}%
\qquad m_{4i}=v_{ix}^{2}+v_{iy}^{2}+\eta _{i}^{2}\text{,}\qquad
m_{5i}=v_{ix}^{2}+v_{iy}^{2}\text{,}\qquad m_{6i}=v_{ix}^{2}-v_{iy}^{2}\text{%
,}\qquad m_{7i}=v_{ix}v_{iy}\text{,}\qquad
m_{8i}=v_{ix}(v_{ix}^{2}+v_{iy}^{2}+\eta _{i}^{2})\text{,}\qquad
m_{9i}=v_{iy}(v_{ix}^{2}+v_{iy}^{2}+\eta _{i}^{2})\text{,} \qquad
m_{10i}=v_{ix}(v_{ix}^{2}+v_{iy}^{2})\text{,}\qquad
m_{11i}=v_{iy}(v_{ix}^{2}+v_{iy}^{2})\text{,}\qquad
m_{12i}=v_{ix}(v_{ix}^{2}-v_{iy}^{2})\text{,}\qquad
m_{13i}=v_{iy}(v_{ix}^{2}-v_{iy}^{2})\text{,}\qquad
m_{14i}=(v_{ix}^{2}+v_{iy}^{2})(v_{ix}^{2}+v_{iy}^{2}+\eta _{i}^{2})\text{,}%
\qquad m_{15i}=v_{ix}v_{iy}(v_{ix}^{2}+v_{iy}^{2}+\eta _{i}^{2})\text{,}%
\qquad m_{16i}=(v_{ix}^{2}-v_{iy}^{2})(v_{ix}^{2}+v_{iy}^{2}+\eta _{i}^{2})%
\text{,}$ where $i=1,\cdots ,16$.

The corresponding equilibrium distribution functions in KMS are as
follows: $\hat{f}_{1}^{eq}=\rho \text{,}\qquad
\hat{f}_{2}^{eq}=j_{x} \text{,}\qquad \hat{f}_{3}^{eq}=j_{y}
\text{,}\qquad \hat{f}_{4}^{eq}=e \text{,}\qquad
\hat{f}_{5}^{eq}=2P+(j_{x}^{2}+j_{y}^{2})/\rho\text{,}\qquad
\hat{f}_{6}^{eq}=(j_{x}^{2}-j_{y}^{2})/\rho \text{,}\qquad \hat{f}%
_{7}^{eq}=j_{x}j_{y}/\rho \text{,}\qquad \hat{f}_{8}^{eq}=(e+2P)j_{x}/\rho
\text{,}\qquad \hat{f}_{9}^{eq}=(e+2P)j_{y}/\rho \text{,}\qquad \hat{f}%
_{10}^{eq}=(4P+j_{x}^{2}/\rho +j_{y}^{2}/\rho )j_{x}/\rho \text{,}\qquad
\hat{f}_{11}^{eq}=(4P+j_{x}^{2}/\rho +j_{y}^{2}/\rho )j_{y}/\rho \text{,}%
\qquad \hat{f}_{12}^{eq}=(2P+j_{x}^{2}/\rho -j_{y}^{2}/\rho )j_{x}/\rho
\text{,}\qquad \hat{f}_{13}^{eq}=(-2P+j_{x}^{2}/\rho -j_{y}^{2}/\rho
)j_{y}/\rho \text{,}\qquad \hat{f}_{14}^{eq}=2(b+2)\rho
R^{2}T^{2}+(6+b)RT(j_{x}^{2}+j_{y}^{2})/\rho +(j_{x}^{2}+j_{y}^{2})^{2}/\rho
^{3}\text{,}\qquad \hat{f}_{15}^{eq}=[(b+4)P+(j_{x}^{2}+j_{y}^{2})/\rho
]j_{x}j_{y}/\rho ^{2}\text{,}\qquad \hat{f}%
_{16}^{eq}=[(b+4)P+(j_{x}^{2}+j_{y}^{2})/\rho
](j_{x}^{2}-j_{y}^{2})/\rho ^{2}$.

\end{document}